\documentclass[manuscript]{acmart}
\usepackage[]{hyperref}

\usepackage{amsmath,amsfonts}
\usepackage{algorithmic}
\usepackage{graphicx}
\usepackage{textcomp}
\usepackage{multirow}
\usepackage{xcolor}
\usepackage{circledsteps}
\usepackage[ruled,linesnumbered]{algorithm2e}
\usepackage{soul}
\usepackage{layouts}

\newcommand{\myparagraph}[1]{\noindent \textbf{#1}.}

\newcommand{\rev}[1]{#1}

\newcommand{\dlop}[1]{{\tt #1}}

\title{Kitsune: Enabling Dataflow Execution on GPUs} 
    
\author{Michael Davies}
\affiliation{ 
    \institution{NVIDIA}
    \country{USA}
}
\email{karus@nvidia.com}

\author{Neal Crago}
\affiliation{ 
    \institution{NVIDIA}
    \country{USA}
}
\email{ncrago@nvidia.com}    
    
\author{Karthikeyan Sankaralingam}
\affiliation{ 
    \institution{NVIDIA}
    \country{USA}
}
\email{karus@nvidia.com}

\author{Stephen W. Keckler}
\affiliation{ 
    \institution{NVIDIA}
    \country{USA}
}
\email{skeckler@nvidia.com}    

\begin{document}

\begin{abstract}
State of art DL models are growing in size and complexity, with many modern models also increasing in heterogeneity of behavior. GPUs are still the dominant platform for DL applications, relying on a bulk-synchronous execution model which has many drawbacks and is ill-suited for the graph structure of DL applications. Many industry and academic works attempt to overcome these by employing vertical fusion but this approach still fails to realize three untapped opportunities: (1) the fact that many resources on the GPU are idle while only one operator executes due to temporal multiplexing of the SM; \rev{(2) lower energy from more intelligent on-chip data-movement which lends to higher performance in a power-provisioned environment.} (3) inability to exploit hidden or reduction dimensions as a source of parallelism to ease pressure on batch size. 
This paper explores relatively uncharted territory, answering the following key question: \emph{Can modest adjustments to the current GPU architecture enable efficient dataflow execution, thereby circumventing the constraints of vertical fusion without necessitating a clean-slate architecture design.}
We develop Kitsune -- a set of primitives that enable dataflow execution on GPUs and an end-to-end compiler based on PyTorch Dynamo. Across 5 challenge applications, Kitsune can provide 1.3$\times$-2.3$\times$ and 1.1$\times$-2.4$\times$ performance improvement as well as 41\%-98\% and 16\%-42\% off-chip traffic reduction for inference and training, respectively.
\end{abstract}

\maketitle

\pagestyle{plain}

\section{Introduction}
\begin{figure}
    \centering
    \includegraphics[width=1.0\columnwidth]{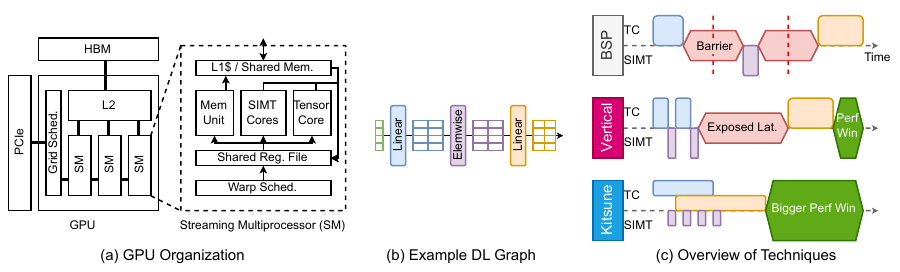}
    \caption{(a) Overview of GPU organization, (b) example DL graph, and (c) stylized comparison of execution techniques. In (c), TensorCore and SIMT resources of the GPU are depicted separately.}
    \label{fig:overview}
\end{figure}

Graphics Processing Units (GPUs) have become the dominant platform for executing deep learning (DL) algorithms due to their amenability to matrix-multiplication and other common DL operations. Historically designed for Single Instruction, Multiple Thread (SIMT) execution with extensive register files, GPUs have evolved significantly. They now boast intricate memory hierarchies, specialized Tensor Cores for general matrix-multiply (GEMM) computations, and support for atomic memory instructions~\cite{hopper-whitepaper}. Depicted in Figure~\ref{fig:overview}, GPUs (a) employ a relatively simple bulk-synchronous programming (BSP) model (c), where a set of independent work items for a single operator (commonly implemented as a single kernel) are run to completion followed by a global barrier before the next set is dispatched. \textbf{However, the BSP model is a misfit to the directed-acyclic graph structure of DL applications, and hence encounters inefficiencies centered around three key areas:} the inability to exploit on-chip data locality of intermediate data passed between operators due to large memory footprints spilling to DRAM, and idle resources due to limited parallelism or low arithmetic intensity within operators.
 




Vertical fusion, depicted in Figure~\ref{fig:overview} (c), is an approach for GPUs to amortize kernel launch overheads and improve data locality between operators and thus reduce off-chip memory traffic through fusing multiple operators into a single CUDA ``mega kernel'', establishing the need for flexibility in GPU execution. Under this paradigm, GPU's execution resources are \emph{temporally} multiplexed between several ``fused'' operators, interleaving partial executions of each operator, allowing tiles of intermediate data to stay resident on chip for reuse, and removing the need for kernel barriers between fused operators. \rev{This multiplexing is depicted in Fig~\ref{fig:overview} (c) by how at a given time, only the TensorCores or SIMT resources are active}. This technique has been commercialized in tools such as TensorRT~\cite{tensorrt} and advanced through academic endeavors like Welder~\cite{shi2023welder}, Astitch~\cite{zheng2022astitch} and others~\cite{huang2023alcop,zhao2022apollo,zheng2023chimera,jung2021deepcuts,zhao2023effectively}. Despite its effectiveness, vertical fusion leaves three performance opportunities untapped. First, because of temporal multiplexing, the technique does not take advantage of the many idle resources available while one operator is executing. \rev{Second, because of how vertically fused operations are structured, spilling large intermediates to DRAM can become unavoidable, incurring a round-trip DRAM latency penalty.} Third, it is unable to exploit reduction or hidden dimensions for parallelism to ease the need for large batch-level parallelism.

Many academic and industry approaches (Groq for e.g.) recognize that dataflow execution (i.e. concurrently executing operators across \emph{space} rather than time) aligns more naturally to the graph structure of DL applications -- mitigating the above inefficiencies of BSP and vertical fusion with \emph{clean-slate} architectures~\cite{shao_simba_2019,lie2023cerebras,vasiljevic_compute_2021,xilinx-versal,prabhakar_sambanova_2022,prabhakar_plasticine_2017,9138986}. The focus of these efforts is dataflow execution of DL (sub)graph nodes at the single-chip level, while recognizing other aspects of dataflow execution also exist at the system level~\cite{barham2022pathways} and within the matrix-engines themselves~\cite{chen2019eyeriss,domke2021matrix,reuther2020survey}. \textbf{This paper explores relatively uncharted territory, answering whether modest adjustments to the dominant GPU architecture and software stack can enable efficient dataflow execution at the chip-level.}







Our key insight is two complementary software-hardware primitives are sufficient to enable dataflow execution on GPUs. They are: 1) a software-only ring queue which facilitates inter-CTA (Cooperative Thread Array) communication by using the L2 cache and global atomics; 2) a modest change to the GPU's grid scheduler to enable it to exploit the heterogeneity of concurrently executing operators. We find that an effective end-to-end compiler can be built that uses these primitives to allow automatic lowering of DL applications to dataflow execution on GPUs, avoiding the need for new IRs or a complex code generation backend. This system, named ``Kitsune'', addresses the problems of the BSP model: executing more than one operator concurrently and passing tiles of intermediate data through on-chip queues increases available parallelism and reduces memory bandwidth pressure. This is depicted in Figure~\ref{fig:overview} (c), \rev{where multiple different operators can executed simultaneously across both the SIMT and TensorCore resources on the GPU}.








The contributions of this work are as follows. 

\begin{enumerate}
    \item \textbf{A systematic characterization of DL applications that highlights the mismatch between graph behavior and GPU bulk-synchronous execution.}

    \item \textbf{A design and analysis of Kitsune's SW/HW primitives needed to enable synchronous dataflow execution on GPUs.}
    
    \item \textbf{A design and implementation for the Kitsune compiler which enables applications to transparently leverage dataflow execution on GPUs.}
    
    \item \textbf{An evaluation of Kitsune across several diverse DL models, spanning inference and training, on a SOTA A100 class GPU.} We show $1.3\times$ to $2.4\times$ speedups, with 16\%-98\% reduction in memory traffic (which indirectly serves as a form of power/energy savings). We also compare Kitsune to SOTA vertical fusion techniques and elucidate the reasons why Kitsune is able to achieve superior performance.
    
    \item \textbf{A sensitivity study of Kitsune's hardware synergy.} When increasing inexpensive hardware resources by 2$\times$ (on-chip compute, on-chip L2 cache bandwidth), while keeping expensive resource (memory bandwidth) unmodified, Kitsune effectively achieves 47\% and 27\% speedup for inference and training, respectively, while baseline execution shows only 18\%-26\%.
\end{enumerate}


\section{Background}\label{sec:bg}
This section presents an overview of Deep Learning, GPU hardware, it's connection to the BSP execution model, and pointers to recent hardware support.

\myparagraph{Deep Learning} 
DL applications use learned parameters to make predictions on data across a variety of application domains, combining input and parameter \textbf{tensors} (multidimensional arrays) with mathematical \textbf{operators} such as linear projections (\dlop{Linear}) to produce outputs. A computational graph is constructed during execution which is then used in automatic-differentiation for computing gradients to ``train'' parameters. Common operators include linear projection, element-wise operators such as ReLU and addition, attention, layernorm, softmax, and convolution. Linear projection, attention, and convolution are all computationally similar; reducing to general matrix-multiplications (GEMMs) whose dimensions are dictated by the operator parameters. Often GEMMs are colloquially used to express the entirety of work done by these operators.

\myparagraph{GPU Hardware}
Figure~\ref{fig:overview} (a) presents a modern GPU chip design~\cite{gpu-background}. A GPU comprises a set of multiple Streaming Multiprocessor (SM) processing cores, a globally shared L2 cache (among all the SMs), and main memory accessible through a high bandwidth interface. SM execution is managed by a GPU-global grid scheduler which is responsible for dispatching work sent from the driver over PCIe. 
The SM includes local data storage, including a large register file and a memory that can either be configured as an L1 cache or a software-managed scratchpad memory (also known as shared memory). Each SM also includes compute functional units for general computation (SIMT Cores), and dedicated  hardware for accelerating tensor operations such as matrix-multiplication (Tensor Cores). The memory system additionally includes support for atomics which are facilitated by the L2. SM counts range from 80 for V100~\cite{v100-whitepaper}, 108 for A100~\cite{a100-whitepaper}, and 132 for H100~\cite{hopper-whitepaper}. Roughly speaking, the L2's bandwidth is $3\times$ of main memory bandwidth~\cite{jia2019dissecting,jia2021dissecting,jia2018dissecting}. 

\myparagraph{GPU Execution model}
A GPU \textbf{kernel} (typically mapped one-to-one with DL operators) is code that is compiled and run on the GPU's SIMT abstraction. Kernels are run with a BSP execution model where one kernel occupies the GPU at a time and finishes completely before the next kernel starts. Each kernel's threads are organized into collections of threads known as \textbf{cooperative thread arrays} (CTAs, a.k.a. ``threadblocks''), and all the CTAs of a kernel make up a \textbf{grid}. A CTA is a non-divisible quanta of work that is mapped to and executes to completion on an SM, where each thread maintains private state in the register file, and communicates with other threads in the same CTA via shared memory.

In the microarchitecture, threads within a CTA are grouped into fixed-size \textbf{warps} (32 for most modern GPUs) which execute in lock step. In modern GPUs, multiple CTAs can run simultaneously on a single SM.  Modern GPUs allow multiple grids to execute simultaneously in limited situations, and have included rich support for atomic memory operations, allowing threads within a CTA, grid and across grids to synchronize with global atomics~\cite{cuda-atomics}. CUDA Streams~\cite{luitjens2015cuda} and CUDA Graphs~\cite{cuda-graphs} are APIs that enable users to specify which kernels are independent and can run simultaneously. 
In practice, neither of these result in co-executing kernels -- due to first-in-first-out ordering and queuing hardware in the global grid scheduler. Current GPUs restrict that a new kernel can only start dispatching once all the CTAs from the current one have dispatched resulting in minimal execution overlap of the two kernels~\cite{cuda-mps, ukarande2021taco}.

\section{Motivation and Program Behavior Characterization}\label{sec:behavior}


\begin{figure}
    \centering
    \includegraphics[width=1.0\columnwidth]{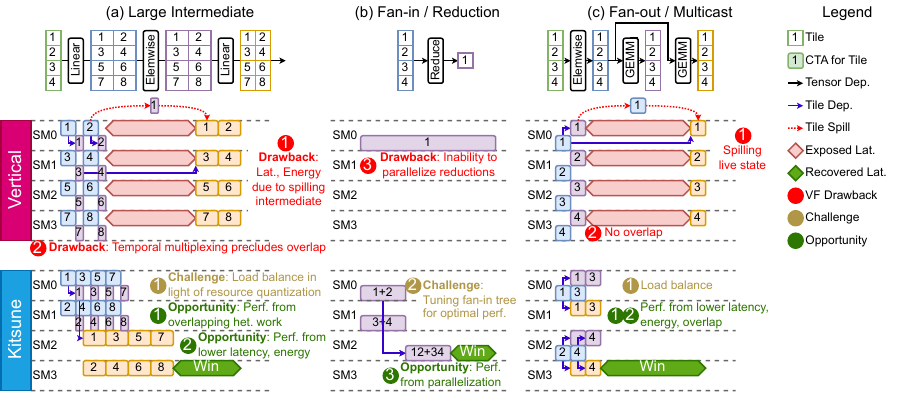}
    \caption{Visualization of the difference between Kitsune and Vertical Fusion for (a) an MLP with a large hidden dimension, (b) a reduction operation, and (c) an operator that sends intermediate to multiple consumers.}
    \label{fig:compare-new}
\end{figure}

In this section, we motivate dataflow execution by examining the opportunities present across a range of DL applications' operator graphs. The applications we focus on are summarized qualitatively in Table~\ref{tab:app-desc}. 
Our DL applications include DLRM~\cite{naumov2019deep}, MeshGraphNets~\cite{pfaff2020learning}, NeRF~\cite{mildenhall2021nerf}, GraphCast~\cite{lam2022graphcast}, and Llama 3 8B~\cite{llama3}. \rev{Note for Llama 3, we discuss it in terms of three separate use-cases: (1) training which encompasses the forward and backward pass for a whole set of tokens, (2) context-phase (``ctx'') which encompasses just the forward pass prefill step of inference, and (3) decode-phase (``tok'') which encompasses the autoregressive token-generation step of inference. The context and decode phases are inference only and will not appear in training results.} We first discuss several common patterns, summarized in Figure~\ref{fig:compare-new}, which we observe are frequently exhibited in popular DL applications, focusing on the limitations of state-of-art vertical fusion compared to Kitsune dataflow.

\begin{table}
    \small
    \centering
    \caption{Description of selected applications. }
    \label{tab:app-desc}
    \begin{tabular}{lrl}
\hline
\textbf{Application} & \multicolumn{1}{l}{\textbf{Year}} & \textbf{Use Case}      \\ \hline
DLRM                 & 2019   & Predicting ad clicks  \\
MeshGraphNets        & 2020   & Mesh based physical simulation     \\
NeRF                 & 2021   & View synthesis     \\
GraphCast            & 2022   & Weather forecast prediction      \\ 
Llama 3 8B           & 2024   & Language modeling \\ \hline
\end{tabular}
    \vspace{-0.1in}
\end{table}

\myparagraph{Operator Patterns}
\rev{
Figure~\ref{fig:compare-new} depicts three common graph patterns composed from \dlop{Linear}, \dlop{Elementwise}, and \dlop{Reduce} operators. These patterns are abstracted from detailed shapes for our applications encompassing examples found in both inference (forward-pass) and training (back-propagation). \dlop{Elementwise} and \dlop{Reduce} operations are not computationally intensive and cannot use the TensorCores on the GPU, unlike GEMM operations which do. Fig~\ref{fig:compare-new} (a) depicts a common scenario where a linear layer (I.e. a GEMM) produces a large output dimension (``N'') which is then fed to a downstream \dlop{Elementwise} and subsequent linear layer. This is seen in many MLPs, and is especially common in the feed-forward network in transformer models which perfrom an projection (linear layer) into a high-dimension followed by a non-linear operation and subsequent projection back to a smaller dimension. Fig~\ref{fig:compare-new} (b) depicts a simple reduction operation. This can be found typically in split-K GEMM operations where partial sums need to be reduced. In addition, reductions over the batch dimension are very common in back-propagation. Finally, Fig~\ref{fig:compare-new} (c) depicts a scenario where one \dlop{Elementwise} feeds two consumer \dlop{GEMM} operations. This is very common in back-propagation, notably for a \dlop{Linear}-\dlop{Activation} pair the backward pass involves computing the gradient for the activatino function which feeds two gradient GEMMs - one for the input activation and one for the weights.
}



\myparagraph{Vertical Fusion Mechanism} 
Vertical fusion seeks to improve DL performance by combining multiple DL nodes and \emph{temporally} switching between partial executions of each node to avoid main-memory traffic of intermediate data. Different code-regions in a single vertically fused kernel encode the entire computation of the fused subgraph, with each CTA working on data-parallel shards of the problem. Keeping with the BSP execution model in which vertical fusion operates, CTAs do not interact with each other and tiles of intermediate data are only passed between the partial executions \emph{within} a CTA. Therefore, implementations of vertical fusion prioritize staging data in shared memory or the register file~\cite{shi2023welder,zheng2022astitch}. 

\begin{figure}
    \includegraphics[width=0.5\columnwidth]{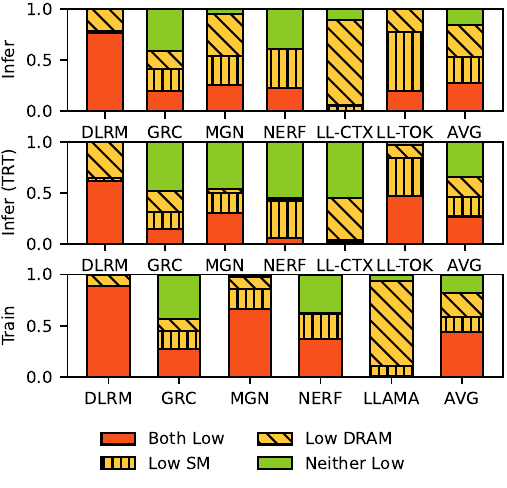}
    \caption{Application runtime spent in different combinations of measured SM and DRAM utilization. Low utilization means less than 33\% of peak.}
    \label{fig:utilbars}
    
\end{figure}

\myparagraph{Vertical Fusion's Utilization Limitations}
Vertical fusion is \textit{unable to exploit idle GPU resources}. Figure~\ref{fig:utilbars} shows, for our application selection, a breakdown of application runtime with respect to SM and DRAM utilization measured from performance counters by NSIGHT Compute for vanilla PyTorch and inference compiled with TensorRT (representative of vertical fusion). We define "low" utilization as less than 33\% of peak, generating four categories. "Both Low" implies that both DRAM and SM utilization are less than 33\%, "Low SM" and "Low DRAM" categories have only one resource below 33\% utilization, and "Neither Low" is time spent with DRAM and SM utilization above 33\%. While "Low" categories indicate portions of time spent with GPU resource severely underutilized, there remains some opportunity even in the "Neither Low" case.




Note that TensorRT does not support training so we only show TensorRT result for inference. Across our applications for bulk-synchronous (unfused) execution, we see 20-25\% and 37-67\% of runtime is spent with both low SM and DRAM utilization for inference and training, respectively; with the exception of DLRM (which has 77\% and 89\%) and Llama Context / Train (which has 0.1\% and 0.3\%). Indeed TensorRT fusion does improve this picture for inference with all applications showing a decrease in ``low'' utilization with the exception of MeshGraphNets. Despite this, there is still ample opportunity for dataflow to capitalize on idle resources shown by the large amount of runtime spent with low utilization of one or both resources. Even if neither resource are low, as exemplified by NERF inference with TensorRT, there's still opportunity for dataflow: operators executing by dataflow eliminates DRAM traffic which would lower the effective DRAM utilization, leading to additional headroom.


\myparagraph{Vertical Fusion's Coverage Limitations} \rev{We discuss coverage limitations by considering graph patterns shown in Figure~\ref{fig:compare-new}. In Figure~\ref{fig:compare-new} (a), when an operator produces an intermediate with a large hidden dimension (E.g. MLP with $N \geq 768$ on an A100 with 192 KB of shared memory), the resultant GEMM tiles exceed the shared-memory capacity. Because of this, even modestly sized intermediates can cause spills to off-chip DRAM\footnote{Indeed the L2 could provide some additional buffering but since every SM runs a data-parallel replica of the same subgraph with the same intermediate storage requirements, that capacity is quickly overrun as well.} As a result of this, the latency from a round-trip to/from off-chip DRAM is incurred for spilled data. On an A100 GPU, this latency is $\approx 409$ns or $572$ cycles at $1.4$GHz. In addition to spilling, because of how vertically fused kernels temporally multiplex the SM, either the SIMT cores or Tensor cores will be idle during computation of each operation, leading to under-utilization of the SM. Naively mitigating this by assigning multiple CTAs to an SM has a major drawback of cutting the effective shared-memory per CTA by the same factor, exacerbating the capacity problem.
}

\rev{
Figure~\ref{fig:compare-new} (b) depicts a reduction operation. One notable and unavoidable place where reductions are common is in back-propagation where gradients are often reduced over the batch dimension. Despite the batch dimension usually being a source of abundant parallelism, here neither BSP or Vertical Fusion are able to extract parallelism from the batch dimension for gradient reduction operation. This means that a small number of CTAs end up performing a reduction, leaving most SMs idle. 
}

\rev{
Finally, Figure~\ref{fig:compare-new} (c) depicts a case where one operator's output is consumed by multiple downstream operations. In particular, this pattern of multicast is representative of back-propagation for a standard \dlop{Linear+Activation} graph. Similar to (a) we find this can lead to spilling tiles of data to off-chip memory since the state needed for one successor child may over-run the shared memory, evicting an intermediate that is needed for a different child. We also see that heterogenous operations cannot simultaneously execute on the SM, leading to underutilization. In general, we observe prior work on vertical fusion does not support back-propagation at all, though we depict in our figure how it would be implemented.
}

\myparagraph{Kitsune}
Our insight is that dataflow -- i.e. having different operators co-execute \textit{spatially across SMs}, rather than temporally switching between executing operators across \emph{time} -- solves all these problems, while preserving the benefits of vertical fusion. Kitsune implements dataflow execution by mapping single operators to CTAs, then passing tiles of intermediate data to downstream operator CTAs using inter-CTA queues residing in on-chip memory to avoid off-chip memory accesses. In doing so, operator CTAs are concurrently mapped and executed across the SMs of the GPU.
Multi-cast and parallel reduction simply become one-to-many and many-to-one communication patterns using our data-queue. The capacity issue, is then trivially solved by splitting hidden dimensions spatially.
Using our modified grid scheduler, hardware under-utilization can be solved by assigning different \textit{types} of CTAs to an SM for co-execution. 

\rev{
Revisiting Figure~\ref{fig:compare-new}, Kitsune can extract performance wins for all of these patterns. First, Kitsune is able to simultaneously execute heterogeneous operations on an SM, addressing under-utilization. Second, with significantly reduced data-movement (especially to/from off-chip DRAM) energy is saved, potentially allowing for higher clock frequencies to be sustained. Finally third, Kitsune is able to extract parallelism from hidden and reduction dimensions.
}

\section{Kitsune Primitives for Dataflow on GPUs}\label{sec:prims}
Kitsune enables the GPU to logically operate with a synchronous dataflow execution model that relaxes the assumptions of bulk-synchronous execution, relying on and leveraging dependence and communication \emph{between} CTAs from different pipeline stages. The execution model comprises of CTAs explicitly communicating with each other which triggers and throttles execution speeds. When data is available in a queue, a CTA starts its execution, writing results to its producer queue. When there is no data in its queue, it idles. The \textbf{first} node of a subgraph reads activations from main-memory (essentially outputs of preceding subgraphs or bulk-synchronous code), and the last node writes results to main-memory. In addition to reading from a queue, a CTA is free to read any other values from memory, and similarly can write to main-memory in addition to writes to its producer queue to trigger its successor. 
In the formal context of execution models models~\cite{1458143}, Kitsune falls under the category of \textit{synchronous dataflow}. Future work can examine further extensions like dynamic dataflow. 

The following subsections develop Kitsune's two key primitives that enable this synchronous dataflow execution model. The first is a synchronized queue structure which allows inter-CTA communication (\S\ref{subsec:queue}). The second is a modified grid scheduler that exploits heterogeneity among executing CTAs to facilitate fine-grained overlapping execution on the SM (\S\ref{subsec:assignment}). We conclude this section by discussing the logical execution model that Kitsune's primitives now provide.

\subsection{Producer consumer communication}\label{subsec:queue}

\begin{figure*}
    \centering
    \includegraphics[width=1.0\textwidth]{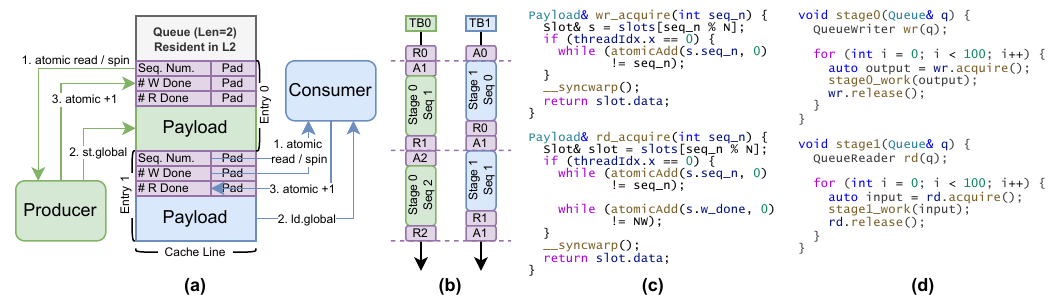}
    \vspace{-0.35in}
    \caption{Queue design. Note: release routines are not shown for space reasons. They involve simple {\tt atomicAdd} calls to update synchronization metadata and a CTA barrier with {\tt \_\_syncthreads()}.} 
    \label{fig:queue-design}
\end{figure*}    

We use GPU atomics to design a synchronized, ring buffer queue for passing data between CTAs. Queues are pinned in the L2 cache using CUDA API functions~\cite{a100-whitepaper} (Fig~\ref{fig:queue-design}(a)). Each entry contains metadata protected by atomic accesses. Figure~\ref{fig:queue-design} shows (a) a diagram of our queue design (with two entries for double-buffering), (b) a timeline of producer-consumer operations, (c) stylized code implementing the queue, and (d) application-level usage. Two CTAs communicating (Fig~\ref{fig:queue-design}(b)) ``acquire'' and ``release'' entries, achieving ordering via sequence numbers. The producer and consumer acquire entries ({\tt wr\_acquire} and {\tt rd\_acquire} in Fig~\ref{fig:queue-design}(c)) by spinning on metadata variables until an entry is freed for use. {\tt acquire} and {\tt release} are exposed as an API which handle sequencing automatically. Typically, only one CTA is spinning on a variable at a given time -- meaning our queue design results in very low contention. 

Our queue is implemented as a library with two API functions: {\tt acquire} and {\tt release}. This allows for easy software integration, introducing minimal overhead exploiting the modern GPU's sophisticated warp-scheduler. Queue code is wrapped with {\tt if threadid==0}, ensuring only one thread in a CTA does any of the queue management. To avoid data-races, ``release'' operations require a CTA-level barrier. Figure~\ref{fig:queue-design}(d) shows how it can be used intuitively by a CUDA programmer or inserted by a source-to-source compiler into existing CUDA kernels. Synchronization variables are all padded to the size of a cache line to avoid false-sharing. 

\begin{figure}[t]
    \centering
    \includegraphics[width=0.5\columnwidth]{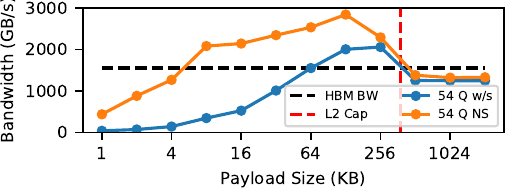}
    \caption{Performance of GPU atomics.} 
    \label{fig:atomics-qperf}
\end{figure}

\myparagraph{Queue Performance}\label{subsec:queue-eval}
Using a microbenchmark, we measure the A100 can sustain 100 M atomics / sec / CTA when under no contention. Based on additional measurements, we find this lends to an upper bound of 385-1541 GB/s \textit{per queue}, far exceeding L2 and HBM bandwidth ($\approx$61 GB/s per SM). 
We evaluate our queue's performance by measuring SM-SM bandwidth with varying payload sizes for 54 queues (108 CTAs for the 108 SMs of the A100 GPU). 
Figure~\ref{fig:atomics-qperf} shows queue management overhead by measuring the performance of data transfers with and without synchronizing atomics enabled.
We find with 128-256 KB payloads, aggregate bandwidth reaches 2 TB/s (37 GB/s/queue). Beyond 256 KB, performance drops due to queue sizes reaching the L2 capacity, causing accesses to spill out to HBM (Limiting us to 1.5 TB/s for A100). Synchronization overhead is large for small queue sizes: 12$\times$ reduction in bandwidth for 1KB payloads. With larger payloads this reduces: synchronization overhead is less than 63\% for $\geq$64KB payloads. 
\textit{Overall, we find GPU global atomics performance is more than enough for our use case. We also find our atomics-based L2 resident queue provides substantial inter-CTA communication bandwidth even in the presence of contention for payloads ranging between 64-256KB.} 



\subsection{Scheduling heterogeneous CTAs}\label{subsec:assignment}
In order to capitalize on idle resources of the SM -- for example, make full use of both the Tensor Core and SIMT Core simultaneously -- we propose a modest change to the CUDA API and GPU Grid Scheduler to specify spatial pipelines (shortly defined) of kernels and maximize GPU resource usage. This is important for enabling and managing true co-execution of kernels which is not supported on current GPUs (\S\ref{sec:bg}). 


\begin{figure}
    \includegraphics[width=0.5\columnwidth]{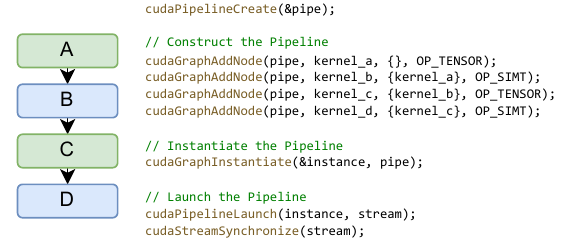}
    \caption{Code snippet of the proposed {\texttt cudaPipeline} API.}
    \label{fig:pipeapi}
\end{figure}

\myparagraph{CUDA API Exposure}
We introduce an abstraction we call a CUDA ``spatial pipeline'', with a similar API to CUDA Graphs but different semantics. Like a CUDA graph, a spatial pipeline specifies a collection of kernels to execute with the key difference being it implies all kernels in the collection require being co-resident on the GPU. The calling code is responsible for limiting the number of CTAs launched per kernel to ensure co-residency is possible (\S\ref{subsec:loadbalancemodel}). Figure~\ref{fig:pipeapi} shows a snippet of host code which specifies and configures the launch of a spatial pipeline. Data dependence information is specified similar to CUDA graphs, and kernels are configured with new metadata that specifies the primary type of dynamic resource they require, either SIMT or TENSOR.

\myparagraph{Hardware Implementation}
To complement our CUDA spatial pipeline abstraction, we propose a modest change to the grid scheduler that allow it to leverage the type information now passed via the kernel call header. On current GPUs, the grid scheduler hardware stores occupancy info for how much of each SM's resources are consumed, which is used to greedily find the first available SM for CTA dispatch using a hardware arbiter (i.e., round-robin)~\cite{ukarande2021taco}. However, this greedy policy doesn't work for Kitsune as it doesn't guarantee overlap; We need to ensure that CTAs of different types are effectively paired for execution on the SM. We augment the round-robin prioritization hardware with two arbiters, one for each type. The two arbiters enables the scheduler to effectively pair different types together, by separately considering dispatch to the same SM. When a new kernel arrives, the arbiter is selected based on the type. The CTA scheduling then proceeds as usual, checking the occupancy of the current SM under consideration for dispatch.

\section{Kitsune Compiler Design}\label{sec:compiler}
In this section we develop the Kitsune compiler, which enables DL applications to transparently leverage dataflow. We implement Kitsune as a PyTorch~\cite{pytorch} compiler backend. We use PyTorch 2.0's Dynamo interface for extracting application graphs including both the forward pass and back-propagation for training. Our compiler backend consumes these graphs and constructs spatial pipelines for execution. 

\begin{figure}
    \includegraphics[width=\textwidth]{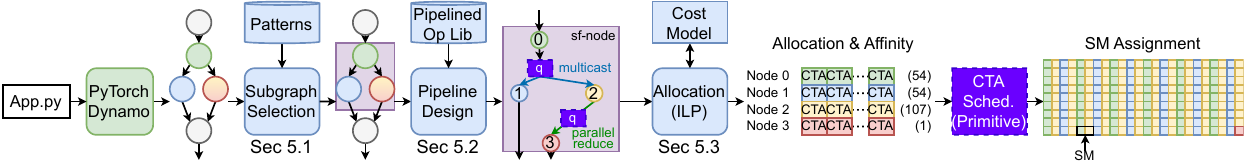}
    \caption{Depiction of the Kitsune compiler flow. Our enabling primitives from the previous section are highlighted in purple.}
    \label{fig:backendflow}
\end{figure}

To realize this compiler, several challenges must be addressed. First, subgraphs must be selected from the original application graph for fusion (\S\ref{subsec:subgraph}). Second, a pipeline must be designed for the subgraph with stages corresponding to operators (\S\ref{subsec:pipedesign}). Third, the stages must be assigned GPU resources to achieve optimal performance addressing load-balancing (\S\ref{subsec:loadbalancemodel}). Figure~\ref{fig:backendflow} depicts how each of these pieces are applied to a PyTorch application. The compiler leverages our software queue structure and modified GPU grid scheduler to enable inter-stage communication and intelligent CTA placement to exploit fine-grain SM resource sharing. Figure~\ref{fig:running-example} depicts how our compiler lowers MeshGraphNets, and will serve as a running example throughout this section.

\subsection{Subgraph Selection}\label{subsec:subgraph}
We first need to select subgraphs for dataflow execution which involves marking groups of DL operators in the computation graph for co-execution in a pipeline. We denote these groups of operators / nodes that form a pipeline as an {\tt sf-node}. The output of this phase is labeled graph with {\tt sf-nodes} identified. At the graph level, a spatially-fused group ({\tt sf-node}) of operations must be ``contiguous'' as defined in~\cite{tarnawski2020efficient} -- that is, there must be no edge which exits the subgraph with a down stream edge that reenters it. Subgraph selection influences pipeline design, allocation, assignment and hence performance, potentially requiring an iterative solution. As a practical solution, we implement a single-pass design that use two rules to exclude a node from a subgraph: nodes that are bulk-sync friendly and nodes that index / gather across all data (gather nodes for embedding for example). With such node exclusions defined, subgraph selection converts to pattern-matching.

Our design and implementation of subgraph selection is heuristic based and uses manual pattern matching. By examining applications properties we identified node patterns that are candidates for subgraph exposing the vulnerabilities of bulk-synchronous execution and vertical fusion. It is essentially a set of regular expressions that express patterns including those seen in Figure~\ref{fig:compare-new}. In particular, our implementation operates at the topological order which linearizes the graph into a list in PyTorch Dynamo (which is deterministic). In practice, additional regular expressions to express different orderings for the topological order are easy. A more formal automata that captures all possible linearizations of a subgraph is beyond the scope of this work. 

We leverage PyTorch's Dynamo to extract whole operator graphs of the forward and backward passes for an application. We then created a library of patterns that expresses patterns that are candidates for subgraphs.
We implement a pattern-matching algorithm for then selecting subgraphs from the original application graph for dataflow execution. This approach searches for user-specified chains of operators in a topological order. Adding new patterns is a trivial task of adding to our pattern library. Figure~\ref{fig:running-example} (a) shows how we selected a subgraph for MeshGraphNets.

\subsection{Pipeline Design}\label{subsec:pipedesign}
The pipeline design problem comprises of inserting queues between nodes of an {\tt sf-node}, and if the work done between two nodes is trivially fusable, fuse them using epilogue fusion (or vertical fusion). The output is a transformed graph which includes one or more queue nodes added, which can then be lowered to CUDA code during code-generation.

\begin{figure}[h!]
    \includegraphics[width=\columnwidth]{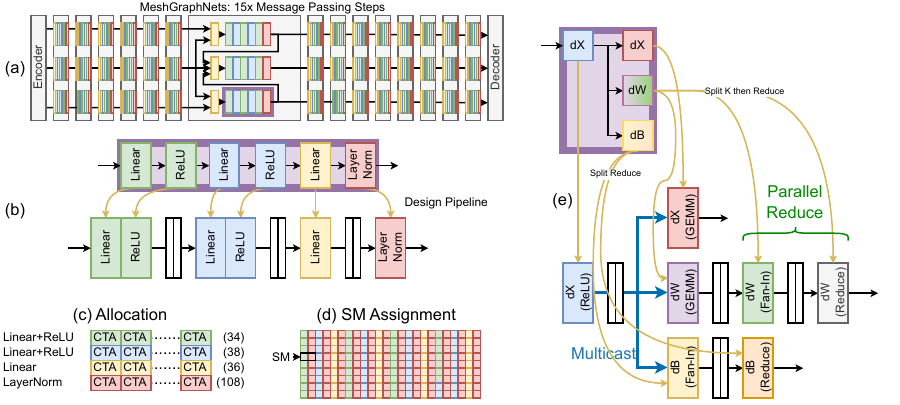}
    
    \caption{Running example of two subgraphs selected from (a) MeshGraphNets. (b) shows an MLP selected from the full application graph forward pass and its pipeline design. (c) and (d) show the allocation and assignment. (e) shows a subgraph from the backward pass for a Linear+ReLU and its pipeline design. We omit allocation / assignment for space. White rectangles in (b) and (e) represent queues.}
    \label{fig:running-example}
    
\end{figure}

{
\begin{algorithm}[t]
\SetAlgoLined
\caption{Algorithm for pipeline design}\label{algo:pipedesignalgo}
\small
\For{$n$ \normalfont{in} $Graph$} {
    \If {\normalfont{IsReduction}$(n)$} {
        $fanin, final \gets$ \normalfont{SplitReduction}$(n)$ \\
        $Graph$.\normalfont{replace}$([n], [fanin, final])$ \\
        $n \gets final$
    }
    \If {\normalfont{IsIntermediate}$(n)$} {
        $q \gets$ \normalfont{CreateQueue}$(n)$ \\
        \For{$c$ \normalfont{in} \normalfont{Dependents}$(n)$} {
            $c$.\normalfont{producer} $\gets q$ 
        }
        $n$.\normalfont{dependents} $\gets [q]$ \\
    }
}

\end{algorithm}
}
Conceptually what this means is taking the original set of operations in the graph and either combining or splitting them to map to pipeline stages that are realized by pipeline-enabled CUDA kernels. For simple patterns \rev{with 1-1 producer-consumer relationships}, the decision is trivial - and involves insertion of queue nodes between nodes of an {\tt sf-node}. For more complex patterns like attention and back-propagation, we implement a parallel reduce which uses our queues to form a reduction tree. Figure~\ref{fig:running-example} (b) and (e) show a pipelined graph starting from our MeshGraphNets subgraph and back-propagation of a single \dlop{Linear} layer, respectively.

In terms of implementation this involves three steps. The graph rewrite algorithm is shown in Algorithm~\ref{algo:pipedesignalgo}. In terms of code-generation, the queue implementation is discussed in \S\ref{subsec:queue}. The third step is to take CUDA kernels and transform them to read/write from queues, instead of from global memory. This last step also includes the process of working on tiles, since a queue's payload needs to be limited. In all cases, the notion of tiling already exists or is trivially doable; for GEMMs the code is already written to work on tiles of inputs and outputs. Completely automating this step for arbitrary code is likely infeasible and involves all the challenges of aliasing analysis etc. For Kitsune, we performed this step manually - it took about 8 person-hours or less for each kernel, with the source-code lines changed ranging from 10 to 40. The limitation this adds to Kitsune is that it is not completely turn-key for new operators not previously seen by the compiler, requiring very modest library modifications of the underlying \textit{new} DL operator. In practice, library developers like NVIDIA and AMD can incorporate such a flow trivially into their development process.

\subsection{Load Balance}\label{subsec:loadbalancemodel}

Load balancing in Kitsune involves the logical allocation of \# CTAs to each node in an {\tt sf-node}. Its output is an allocation as shown in Figure~\ref{fig:running-example}(c). This needs to be done cognizant of overlapped execution of dissimilar CTAs on the same SM.

We use a zero-latency performance model to estimate the throughput of a spatially-fused subgraph based on an allocation of CTAs to each stage. We then formulate the allocation problem as an integer linear program (ILP) which can be used with standard solvers to produce an optimal assignment which maximizes throughput of the subgraph. We augment our ILP formulation to enable over-subscribing CTAs onto SMs to enable overlapping dissimilar behavior -- specifically, we consider two classes of operations: SIMT-heavy, and TensorCore-heavy, and assume an SM can simultaneously execute one of each with no performance degradation. We discuss in \S\ref{subsec:assignment} low level details of how this overlap can be leveraged on modern GPU hardware.

Algorithm~\ref{algo:ilpbalance} shows our ILP formulation. We model throughput as the minimum throughput pipeline stage in the fused subgraph additionally constrained by memory bandwidth and aggregate L2 bandwidth based on analytic evaluation of the total number of bytes read/written from DRAM (DRAM Bytes), and L2 (L2 Bytes). For the $i^{th}$ of $n$ operators in a spatial pipeline, we estimate the performance by combining a measured BSP throughput ($t_i$) with an estimate of how the performance will scale (speedup or slowdown) based on how many CTAs it is assigned ($r_i=$ResourceScale($a_i$)) and an estimate of speedup afforded by operating where some number of its operands are now resident in on-chip storage instead of DRAM ($s_i=$Speedup($a_i$)). In practice, we define Speedup($a_i$) to be $1/u$ where $u$ is the maximum resource utilization of the SIMT or TensorCore pipelines. We allow the number of SIMT and Tensor stages to independently be assigned SMs to exploit overlapping these dynamic resources. In practical deployment terms, we require either a two-pass compiler, run-time optimization pass, or a dictionary of kernel characteristics to get $u_i$ to guide the ILP. Since DL models generally run in a curated environment (TensorRT for example), any of those approaches are practical, and don't introduce any application slowdowns.

\LinesNotNumbered{
\begin{algorithm}[t]
\small
\begin{equation*}
\begin{array}{ll@{}ll}
\text{maximize}     & thrpt & \\[2pt]
\text{subject to}   & thrpt < r_i * s_i * t_i &  & (i=1 ,\dots, n)\\[2pt]
                    & \multicolumn{3}{l}{thrpt * \text{(DRAM Bytes)} < \text{DRAM}_{peak}}\\[2pt]
                    & \multicolumn{3}{l}{thrpt * \text{(L2 Bytes)} < \text{L2}_{peak}}\\[2pt]
                    & \multicolumn{3}{l}{t_i = \text{Bulk-Sync Thrpt. for Op $i$}} \\[2pt]
                    & \multicolumn{3}{l}{r_i = \text{ResourceScale}(a_i)} \\[2pt]
                    & \multicolumn{3}{l}{s_i = \text{Speedup}(a_i)} \\[2pt]
                    & 1 \leq a_i \leq \text{\# SMs} \\[2pt] 
  & 
    \multicolumn{3}{l}{
        \displaystyle\sum\limits_{i=1}^n \text{IsSimt}_i * a_i = \text{\# SMs}
    } \\[2pt] 
                     & 
    \multicolumn{3}{l}{
        \displaystyle\sum\limits_{i=1}^n \text{IsTensor}_i * a_i = \text{\# SMs}
    }
\end{array}
\end{equation*}
    \caption{ILP formulation for load balancing.}
    \label{algo:ilpbalance}

\end{algorithm}
\vspace{-0.15in}

}

\section{Evaluation}\label{sec:eval}
We now examine the effectiveness of Kitsune across our applications and GPU models. We guide our evaluation with the following questions: i) How well does Kitsune support composing arbitrary operations across DL applications? ii) What is the end-to-end performance of applications running with Kitsune and what are the reasons for variation across applications and modes of operation? iii) What is the sensitivity of our performance and gains to machine parameters including on-chip compute (number of SMs), off-chip DRAM bandwidth, and L2 and crossbar bandwidth?

\subsection{Methodology}
Our evaluation is based on running our 5 applications in a validated GPU simulator emulating an A100 GPU which takes as input the compiled versions of our applications. We built our compiler and a queue library (\S\ref{subsec:queue-eval}) (characterized and run on silicon). Because we need our grid scheduler modifications to allow the overlap afforded by Kitsune (\S\ref{subsec:assignment}), we evaluate Kitsune using a modified version of NVIDIA's NVArchSim (NVAS), a hybrid trace- and execution-driven GPU simulator~\cite{villa2021need} that has been validated against NVIDIA's Ampere GPU.
This also allows us to study sensitivity to individual hardware features, instead of being restricted to particular SKUs.

\begin{figure}
    \includegraphics[width=\columnwidth]{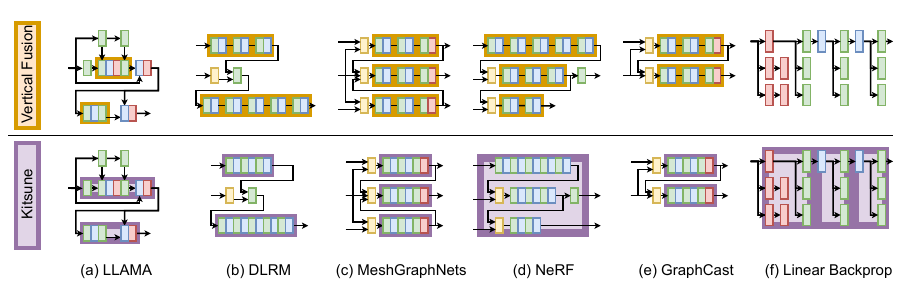}
    \includegraphics[width=2in]{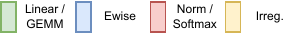}
    \caption{Depiction of applications and the fusions we apply.}
    \label{fig:apps}
\end{figure}

Our baseline for speedup results is unmodified PyTorch execution. We use our compiler and modeling flow based on NVAS to present speedups afforded by both vertical fusion and Kitsune. Figure~\ref{fig:apps} shows the fusions that are chosen by our compiler: thick orange boxes on the left side show the fusions we select based on vertical fusion techniques, while thick purple boxes show the fusions made possible with Kitsune. \textbf{Note: our model of vertical fusion combines the techniques and mechanisms from state-of-art industry and academic approaches of TensorRT~\cite{tensorrt}, AStitch~\cite{zheng2022astitch} and Welder~\cite{shi2023welder}.}

We first describe the quantitative scope of the opportunity that Kitsune provides. We then discuss inference and training separately. For Kitsune, we present results for both the subgraphs of the applications as well as the speedup for the entire application.



\subsection{DL Application Operator Coverage}
Table~\ref{tab:app-sum} provides a characterization of the applications at the DL operator level denoting the number of operators that are grouped into pipelines. The top half of rows are for inference and the bottom half are training. Note this data is for operator count (we discuss time below). For the majority of our applications, $>$70\% of operators are candidates for grouping, with higher coverage for inference. We note that vertical fusion covers only the forward pass operators for training\footnote{We note that none of the academic work or TensorRT have demonstrated execution of training yet - our results are thus optimistic for vertical fusion.} and it's coverage is typically lower.
The last two columns show memory traffic savings both for Vertical Fusion and Kitsune. Traffic savings is useful in itself, as it results in energy/power savings (by downclocking the memory frequency to sustain the lower bandwidth needs). O'Connor et al.~\cite{8686544} and others~\cite{kandiah2021accelwattch,gao2020estimating} have argued that GPUs are becoming memory power limited.



\begin{table}
    \small
    \centering
    \caption{Summary of fusions and traffic reductions.}
    \label{tab:app-sum}
    \begin{tabular}{lrrrrr}
\textbf{}    & \textbf{}       & \multicolumn{2}{c}{\textbf{Fusion Coverage}} & \multicolumn{2}{c}{\textbf{Traffic Red.}} \\
\textbf{App} & \textbf{\# Ops} & \textbf{Vertical}      & \textbf{Kitsune}     & \textbf{Vert.}         & \textbf{Kitsu.}        \\ \hline
\multicolumn{6}{c}{\textbf{Inference}}                                                                                              \\ \hline
DLRM         & 21              & 17 (81\%)             & 17 (81\%)            & 22.53 \%                 & 44.27 \%             \\
GRC          & 35              & 21 (60\%)             & 29 (83\%)            & 23.98 \%                 & 57.20 \%             \\
MGN          & 51              & 36 (71\%)             & 41 (80\%)            & 56.54 \%                 & 57.76 \%             \\
NERF         & 24              & 18 (75\%)             & 24 (100\%)           & 40.19 \%                 & 98.58 \%             \\ 
LL-CTX       & 27              & 10 (37 \%)            & 19 (70 \%)           & 10.04 \%                 & 49.07 \%             \\ 
LL-TOK       & 27              & 10 (37 \%)            & 19 (70 \%)           & 0.01 \%                  & 0.07 \%                \\ 
\hline
\multicolumn{6}{c}{\textbf{Training}}                                                                                               \\ \hline
DLRM         & 59              & 18 (31\%)             & 46 (78\%)            & 7.86 \%                     & 25.07 \%                \\
GRC          & 101             & 20 (20\%)             & 76 (75\%)            & 9.06 \%                     & 40.06 \%                \\
MGN          & 148             & 36 (24\%)             & 108 (73\%)           & 21.76 \%                    & 40.26 \%                \\
NERF         & 69              & 18 (26\%)             & 56 (81\%)            & 14.13 \%                    & 45.47 \%                \\ 
LLAMA        & 88              & 10 (11 \%)            & 34 (39 \%)           & 2.85 \%                     & 45.16 \%                \\ 
\hline
\end{tabular}
    \vspace{-0.1in}
\end{table}

\subsection{Inference Performance}

\begin{figure}
    \centering
    \includegraphics[width=0.5\columnwidth]{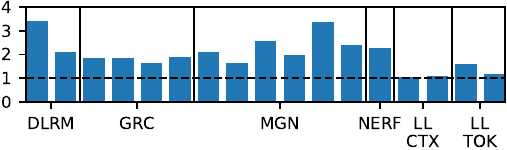} 
    \caption{Inference subgraph speedups including sensitivity to hardware resources.}
    \label{fig:inf-subgraphs}
\end{figure}

\begin{figure}
    \centering
    \hspace{0.5in}
    \includegraphics[width=0.4\columnwidth]{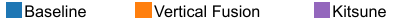} \\
    \includegraphics[width=0.5\columnwidth]{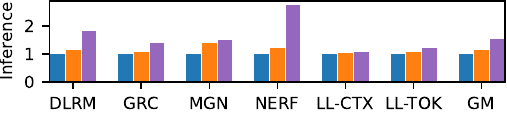} \\
    \vspace{0.1in}
    \includegraphics[width=0.5\columnwidth]{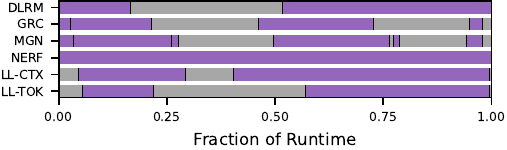} \\
    \vspace{-0.1in}
    \caption{Inference End-to-end Speedup over Bulk-Sync.}
    \label{fig:inf-perf-tl}\vspace{-0.1in}
\end{figure}

Figure~\ref{fig:inf-subgraphs} shows the speedup Kitsune provides for each of the subgraphs in each of the applications. Figure~\ref{fig:inf-perf-tl}'s timeline show the time contributed to overall execution by each of the subgraphs, and in gray we show the time the application spends in kernels/operators that run in bulk-synchronous mode. Figure~\ref{fig:inf-perf-tl}'s bar-charts show full application speedup.

\begin{figure*}
    \centering
    \includegraphics[width=1\textwidth]{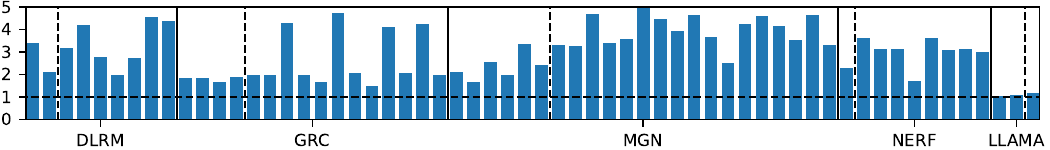} 
    \vspace{-0.3in}
    \caption{Training subgraph speedups including sensitivity. Dashed lines separate forward and backward passes.}
    \label{fig:train-subgraphs}
\end{figure*}   
Overall, sub-graphs speedup range from 1.04$\times$-3.4$\times$ across the applications, with a geomean of 1.9$\times$. The least speedups are for the subgraphs of Llama-Ctx because they are already achieving >50\% of machine peak compute and so do not benefit a lot from operating in spatial mode. 
NeRF is an example where large speedup is achieved (2.3$\times$), highlighting many of Kitsune's benefits: all the nodes of NeRF's forward pass are spatially fused, allowing most layers to pull intermediates from a queue instead of main-memory; and the concat operations are free to occupy the SIMT units of the SMs while the GEMMs use the TensorCores. Due to the intermediate sizes, vertical fusion cannot fuse NeRF's linear layers\footnote{We use the original NERF configuration which uses hidden dim $=$ 256.}. 


When looking at full application performance, we observe two phenomenon: large portions of time are spent in the sub-graphs (typically $>50\%$), and a single application has few subgraphs (the black lines in Figure~\ref{fig:inf-perf-tl} indicate end of a sub-graph). For end-to-end performance, we see geomean $1.5\times$ speedup. Llama-Ctx shows the least speedups because its subgraphs' speedup is modest (4\% - 8\%), despite its sub-graph coverage in time is $84\%$. 

\emph{\textbf{Takeaway:} 
We find Kitsune provides substantial performance opportunity for DL inference with this generally scaling with number of fused operations. We observe DRAM traffic is substantially reduced, suggesting higher performance could be possible without increasing bandwidth.}

\subsection{Training Performance}





Figures~\ref{fig:train-subgraphs} and~\ref{fig:train-perf-tl} show the corresponding results for training, with training broken down further in terms of the forward and backward pass. The forward pass is similar to inference, with the added issue of intermediate activations being stored to main-memory for computing gradients. The backward pass then uses these to compute gradients for parameters.


Considering end-to-end speedup, we see two trends. As expected, the backward pass takes about 2$\times$ the time of the forward pass. Less fractional time of the backward pass is spent in spatial mode, especially for DLRM, where the backward pass for the feature interaction which is not spatially fused takes substantial runtime, causing an Amdahl's law effect on training back-backpropagation. End-to-end speedups range from only $1.1\times$ to as high as $2.2\times$.

\emph{\textbf{Takeaway:} Kitsune still enables performance gains for Deep Learning training, with lower improvements due to smaller fusions in the backward pass compared to forward. Because of Kitsune's ability to parallelize reductions, training benefits more from spatial fusion compared to the parallelism-limited bulk-synchronous baseline.}

\subsection{Comparing to Vertical Fusion}
\begin{figure}[t]
    \centering
    \hspace{0.5in}
    \includegraphics[width=0.4\columnwidth]{figs/e2e-legend.pdf} \\
    \includegraphics[width=0.5\columnwidth]{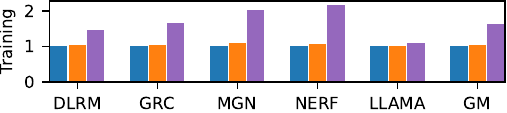} \\
    \vspace{0.1in}
    \includegraphics[width=0.5\columnwidth]{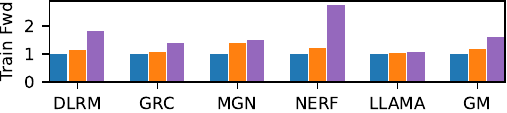} \\
    \vspace{0.1in}
    \includegraphics[width=0.5\columnwidth]{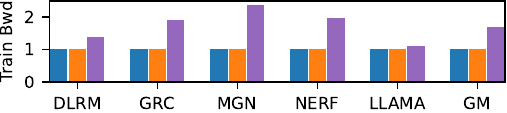} \\
    \vspace{0.1in}
    \includegraphics[width=0.5\columnwidth]{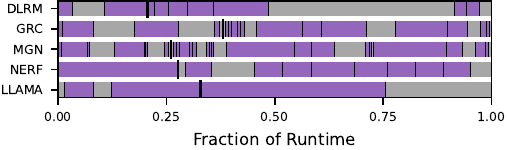} \\
    \vspace{-0.1in}
    \caption{Training End-to-end Speedup over Bulk-Sync.}
    \label{fig:train-perf-tl}
    \vspace{-0.2in}
\end{figure}

Due to the limitations outlined in \S\ref{sec:behavior}, effectiveness of Vertical Fusion is substantially lower than Kitsune for inference, with MeshGraphNets showing the best speedup ($1.4\times$) with geo-mean $1.14\times$ (Figure~\ref{fig:inf-perf-tl}).  Since it only applies for the forward pass, training speedups are even lower (Figure~\ref{fig:train-perf-tl}). Related works like Welder, for inference, have reached similar findings: when applied to production settings of running with TensorCore and meaningful batch-size (like 32 or larger), speedups over un-optimized PyTorch (worse than our baseline) is 30\% or so, with no speedup over TensorRT on Nvidia V100~\cite{shi2023welder}. Those works target additional scenarios like FP32 based computation (thus eliding our overlap opportunity) and edge-case scenarios like batch-size=1, which are less important in production data-center deployment. 
Philosophically they target improvements through software in the configuration space where GPUs are inefficient (bs=1, fp32 mode etc). We focus on production scenarios: batched training and inference using TensorCores to address inefficiencies.

\subsection{Comparing SM and DRAM Utilization}
\begin{figure}
    \includegraphics[width=0.5\columnwidth]{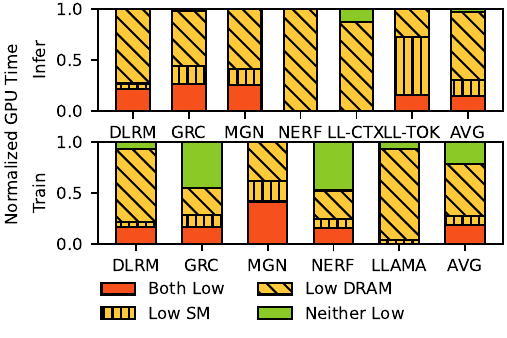}
    \caption{Application runtime spent in different combinations of SM and DRAM utilization as reported by our model. Low utilization means less than 33\% of peak.}
    \label{fig:utilbars_ours}
\end{figure}

Figure~\ref{fig:utilbars_ours} shows a breakdown of application runtime spent with different resource utilization when running with Kitsune. For inference, comparing to our data in Figure~\ref{fig:utilbars}, we see 26\% and 15\% of runtime is spent with both low utilization for BSP and Kitsune, respectively. 
For training, we observe on average, Kitsune only spends 18\% of runtime in low utilization compared to 44\% for bulk-synchronous. In addition, Kitsune on average spends much more runtime with just low DRAM utilization for training: 50\% vs 23\%. This difference is less pronounced for training compared to inference because training requires more DRAM traffic to save intermediate activations for back-propagation. 

\emph{\textbf{Takeaway:} We find Kitsune is able to capitalize on the under-utilized resources of the GPU, reducing runtime spent with low resource utilization for tmost of our applications.}

\section{Related Work}\label{sec:relwork}

\myparagraph{DL Operator Mapping} Pipeline design for Kitsune is related to the problem of ``operator mapping''. This has largely been looked at in the context of spatially exposed hardware for \textit{single operators} including works such as TimeLoop~\cite{parashar2019timeloop}, MAESTRO~\cite{maestro}, AMOS~\cite{zheng2022amos}, and CoSA~\cite{huang_cosa_2021}, which treat an operator as a transformable loop-nest, and TVM~\cite{chen2018tvm} which lowers semantics expressed with einsums to low-level code. 

\myparagraph{DL Operator Fusion} Traditional GPU kernel fusion focuses on fusing memory-intensive kernels together~\cite{qiao2018automatic, wahib2014scalable,wu2012kernel,qiao2019loop}, and modern DL compilers often support simple operator fusion at the register level~\cite{zheng2020ansor, dnnfusion:pldi2021, RAMMER:OSDI2020} or for improving data reuse for identical and related operators~\cite{wang2020accelerating,sivathanu2019astra,jia2019taso}.
Building on single-operator mapping, many recent academic works address vertical fusion including ALCOP~\cite{huang2023alcop}, Apollo~\cite{zhao2022apollo}, AStitch~\cite{zheng2022astitch},   Chimera~\cite{zheng2023chimera}, Deepcuts~\cite{jung2021deepcuts}, GraphTurbo~\cite{zhao2023effectively}, and Welder~\cite{shi2023welder}. We discuss the capability of AStitch, Welder, and state of art vertical fusion in Section~\ref{sec:behavior}. AStitch, Welder and GraphTurbo all use some notion of an anchor-and-propa-gate scheme to handle streaming compatibility between fused layers. Kitsune is more composable and general than all of these, being able to fuse many more operators into co-resident GPU kernels. Other drawbacks and limitations of vertical fusion have been discussed at length in Section~\ref{sec:behavior}.

\myparagraph{GPU Multitasking}
HFuse~\cite{horizontal-fusion:cgo2022} presents a methodology for horizontal fusion which can leverage overlap of heterogeneous work but is restricted to only fusing pairs of nodes with no data dependencies. Works such as ISPA~\cite{zhao2022ispa} and SMK~\cite{wang2016simultaneous} provide a pure software, and hardware-codesign solutions (respectively) for achieving fine-grained multitasking on GPUs. SMK uses hardware mechanisms to enable preemption of CTAs on the SM for ``partial context switching'' -- the goal of which is achieve higher overall utilization of SM resources with heterogeneous CTAs. ISPA uses a pure software approach for co-scheduling pairs of Tensor-heavy and SIMT-heavy kernels. It uses several software techniques to promote efficiency of co-occupancy, but ultimately relies on the existing GPU thread scheduler to make CTA placement decisions. All these approaches focus on co-scheduling just two kernels with no data dependence. Kitsune enables any number of kernels to co-execute in spatial pipelines with data-dependencies supported by our queues and relying on a modified CTA scheduler to make smart decisions about placement of CTAs to best utilize SM resources.

\myparagraph{Data-Triggered Execution} \rev{WorkGraphs~\cite{workgraphs} is a recent development in the graphics space to afford data triggered execution on GPUs. However, it does not address on-chip data-orchestration to maintain cache residency of intermediates. Additionally, it operates on a level of granularity much smaller than Kitsune, using individual records and shader invocations as the unit of work. Kitsune in contrast is designed to orchestrate producer-consumer communication on-chip at a granularity of tensor tiles of around 64KB payloads. Finally, WorkGraphs doesn't support join operations with different input record types, vastly reducing the generality and applicability beyond shader pipelines.}


\section{Conclusion}\label{sec:conc}
We observe that the GPU BSP model limits its effectiveness for various important DL workloads, with state-of-art vertical fusion still leaving performance opportunities untapped. We design and implement Kitsune which enables synchronous dataflow execution for modern GPUs, leveraging existing support for synchronization and integrating into both CUDA and PyTorch. It's only hardware modification is extension of the GPU grid scheduler to be aware of affinity of CTAs to the SIMT vs TensorCore units. Kitsune reduces both main memory traffic and end-to-end runtime across DL networks on GPUs for both inference and training. 

\bibliographystyle{plain}
\bibliography{references}

\end{document}